\documentclass[useAMS,usenatbib]{mn2e}
\usepackage{epsfig}
\usepackage{amsmath}
\newcommand{\be}{\begin{equation}}
\newcommand{\beq}{\begin{equation}}
\newcommand{\ba}{\begin{eqnarray}}
\newcommand{\ee}{\end{equation}}
\newcommand{\eeq}{\end{equation}}
\newcommand{\ea}{\end{eqnarray}}
\newcommand{\msun}{$M_{\odot}\hspace{1mm}$}

\newcommand{\kms}{km s$^{-1}\hspace{1mm}$}
\newcommand{\cm}{cm$^{2}\hspace{1mm}$}
\newcommand{\GHz}{{\rm GHz}}

\newcommand{\kpc}{{\rm kpc}}

\newcommand{\apj}{ApJ}

\newcommand{\apjl}{ApJL}
\newcommand{\mnras}{MNRAS}

\newcommand{\procspie}{Proc. SPIE}

\newcommand{\araa}{ARA\&A}
\def\lsim{~\rlap{$<$}{\lower 1.0ex\hbox{$\sim$}}}
\def\gsim{~\rlap{$>$}{\lower 1.0ex\hbox{$\sim$}}}
\begin{document}

\title[CMB Anisotropies From LBGs]{CMB Anisotropies from Outflows in Lyman Break Galaxies}

\author[Daniel Babich and Abraham Loeb]
{Daniel Babich$^{1,2}$\thanks{E-mail:dbabich@cfa.harvard.edu} and Abraham
Loeb$^{1}$\thanks{E-mail:aloeb@cfa.harvard.edu}\\ $^1$
Astronomy Department, Harvard University, 60 Garden Street, Cambridge, MA
02138, USA\\
$^2$ Current address: California Institute of Technology, Theoretical Astrophysics, MC 130-33
Pasadena, CA 91125, USA}

\date{\today}
\pagerange{\pageref{firstpage}--\pageref{lastpage}} \pubyear{2006}
\maketitle
\label{firstpage}
\begin{abstract}
Thomson scattering of the Cosmic Microwave Background (CMB) on moving
electrons in the outflows of Lyman Break Galaxies (LBGs) at redshifts 2--8
contributes to the small-scale CMB anisotropies. The net effect produced by
each outflow depends on its level of deviation from spherical symmetry,
caused either by an anisotropic energy injection from the nuclear starburst
or quasar activity, or by an inhomogeneous intergalactic environment.  We
find that for plausible outflow parameters consistent with spectroscopic
observations of LBGs, the induced CMB anisotropies on arcminute scales
reach up to $\sim 1 \mu$K, comparable to the level produced during the
epoch of reionization.
\end{abstract}


\section{Introduction}
Several experiments to observe the Cosmic Microwave Background (CMB)
anisotropies on arcminute scales are currently, or will soon be, underway
\citep{Kosowsky03,Ruhl04, Lo05}.  These experiments plan to measure the CMB
power-spectrum for a spherical harmonic multipole index of $10^3 \le \ell
\le 10^4$ at several frequencies centred around the thermal
Sunyaev-Zel'dovich null of 217 \GHz. Photon diffusion damps the CMB
anisotropies on these small scales during cosmological recombination at
redshift $z\sim 10^3$ \citep{Silk68}, and so any observed signal must
originate at much lower redshifts.  Indeed, the above experiments plan to
constrain the epoch of reionization and the growth of structure in the low
redshift universe \citep{Zahn05}.

The primary physical mechanism which is responsible for the small scale CMB
anisotropies is Thomson scattering of CMB photons off moving electrons.
Any peculiar velocity induces a Doppler anisotropy of the scattered
radiation along the direction of motion.  This accounts for the CMB
anisotropies produced by the peculiar velocities of clusters (the so-called
{\it kinetic Sunyaev Zel'dovich effect}) \citep{Zeldovich80}, by peculiar
velocities of linear overdensities in the intergalactic medium (the
so-called {\it Ostriker-Vishniac effect}) \citep{Ostriker86, Vishniac87}
and by the peculiar velocities of the fluctuations in the ionization
fraction during patchy reionization \citep{Gruzinov98}.

In this {\it Letter} we examine the contribution of outflows in Lyman Break
Galaxies (LBGs) to the small-scale CMB anisotropies. LBGs are believed to
be the ancestors of present-day luminous elliptical galaxies. They are
observed to produce gas outflows with velocities of several hundred \kms
(see \cite{Giavalisco02} for a comprehensive review). In contrast with the
traditional kinetic Sunyaev-Zel'dovich effect where the bulk velocity of
the virialized gas is responsible for the induced CMB anisotropy, we focus
here on the Doppler effect of the outflowing gas and ignore any bulk motion
of the LBG as a whole (which produced a smaller effect at the redshifts of
interest). This bulk effect is included in standard calculations of the
non-linear generalization of the Ostiker-Vishniac effect \citep{Hu00}.

The contribution from a single LBG to the fractional temperature
fluctuation of the CMB can be expressed as \be \label{los_int} \frac{\Delta
T}{T} = - \int dl \sigma_T n_e\frac{\vec{n} \cdot \vec{v}}{c}, \ee where
$\sigma_T = 6.65 \times 10^{-25}$ \cm is the Thomson cross section, $n_e$
is the electron number denisty, $\vec{v}$ is the electron peculiar
velocity, $c$ is the speed of light and $\hat{n}$ is the observer's
line-of-sight toward the LBG \citep{Zeldovich80}. The integration traces
the photon's path through the LBG outflow.

The radial extent of the outflow is found by solving the corresponding
hydrodynamics equations. These coupled non-linear partial differential
equations can be reduced to a single ordinary differential equation
\citep{Tegmark93, Furlanetto03} under the assumption that the gas swept-up
by the outgoing blast wave lies in a thin shell behind the propagating
shock front (the so-called {\it thin shell approximation}). The validity of
this approximation is illustrated by the self-similar Sedov-Taylor-von
Neumann solution for a point explosion in which $90\%$ of the swept-up mass
resides in a shell of thickness $10\%$ of the outflow's radius (see
Ostriker \& McKee 1988 and Ikeuchi et al. 1983 for additional discussion on
this approximation).  The thin shell approximation allows us to treat the
radiative transfer of CMB photons through the shock front in a plane
parallel geometry. When the thickness of the shock front is small compared
with the shock front's radius of curvature, the path length through the
shock front can be expressed as \be \delta l \approx \frac{\delta R}{|
\hat{n}\cdot \hat{v} |}, \ee where $\delta R$ is the thickness of the shock
front. Using conservation of mass and the density compression ratio for a
strong adiabatic shock, one gets $\delta R/R = (\gamma -1)/(3\gamma + 3)$,
where $R$ is the radius of the outflow and $\gamma$ is the adiabatic index
of the gas \citep{Ostriker88}.

Within the thin shell approximation the line-of-sight integration in
Eq. (\ref{los_int}) is simplified to \be \left(\frac{\Delta T}{T}
\right)_i = - \frac{\sigma_T~\delta R}{c}~[n_e(\vec{r}_1) v(\vec{r}_1) -
n_e(\vec{r}_2) v(\vec{r}_2)], \ee where $\vec{r}_1$ and $\vec{r}_2$ are the
location, in a coordinate system centred on the LBG, on the shock front 
where the line-of-sight respectively enters and exits the shock front. 
The subscript $i$ labels the contribution from a particular LBG.

If the outflow is spherically symmetric then there is an exact cancellation
of the anisotropies produced at the entry and exit points of the
line-of-sight through the LBG shock front\footnote{Note that even if the
outflow had a perfect spherical shape, the finite light crossing time
through the outflow would produce a net non-zero signal because the flow
parameters are time-dependent.  This effect would produce a signal
$\sigma_T n_e R (\Delta v)/c \sim \sigma_T n_e (dv/dt) (R/c)^2$ that is
extremely small and is ignored here.}. However, observations of
low-redshift starburst galaxies, which serve as analogs of the higher
redshift LBGs, show evidence for highly non-spherical outflow geometries
\citep{Martin99}. It is therefore reasonable to expect that the early
stages of LBG outflows, whether they are driven by starburst or quasar
activity\footnote{In this {\it Letter} we will only include the feedback
driven by supernovae. Energy input from quasars will lead to enhancements
in the bubble size and outflow velocity [see \cite{Furlanetto01} for a
description of quasar outflows].}, would produce CMB anisotropies.

For simplicity, we will assume that the outflow is axisymmetric about some
axis $\hat{z}$ (possibly the rotation axis of the galactic disc or possibly
the jet axis of a central quasar) and expand both the outflow velocity and
shock front density in Legrendre polynomials, $ \vec{v}(\vec{r}) =
\hat{r}~\sum_{\ell} v_{\ell}(r) P_{\ell}(\hat{z} \cdot \hat{r})$ and
$n_e(\vec{r}) = \sum_{\ell} n_{\ell}(r) P_{\ell}(\hat{z} \cdot \hat{r})$,
where we have assumed that the velocity field is radial ($\hat{v} \equiv
\hat{r}$).  Even if the outflow began highly collimated, it will eventually
isotropize as it propagates into the intergalactic medium (IGM; an
analogous tendency exists in relativistic flows, see Ayal \& Piran
2001). On longer timescales, IGM inhomogeneities, such as filaments and
voids, will once again make the outflows non-spherical. Numerical
simulations are needed in order to properly model these effects.  
Here we parameterize the level of asphericity in the outflow by a
coefficient $\epsilon$ (see Eq. \ref{eq:dT2} below).

The outline of this {\it Letter} is as follows.  In \S 2 we analyze the
CMB anisotropy induced by a single LBG with an arbitrarily non-spherical
outflow. In \S 3 we calculate the resulting CMB power spectrum. In \S 4
we present numerical results and in \S 5 we summarize our conclusions.
Throughout our discussion, we will assume the WMAP3 cosmological model
\citep{Spergel06}.

\section{Single LBG Signal}
The signal along a given line-of-sight includes contributions from LBGs
with different masses, outflow ages, orientations of the outflow symmetry
axis and line-of-sight impact parameters. For simplicity, we begin by
analyzing the expectation value for LBGs formed at a certain redshift and
of certain mass and age. Since the net effect from a single LBG can be
either positive or negative we will find that the mean of the signal along
a given direction is always zero, but a non-zero variance will be produced
due to Poisson fluctuations in a manner equivalent to a random walk. The
analysis will be done separately for the distinct cases where either one or
two lines of sight intersect the same LBG outflow.

\subsection{One Sightline}
\begin{figure}
\includegraphics[width=8cm]{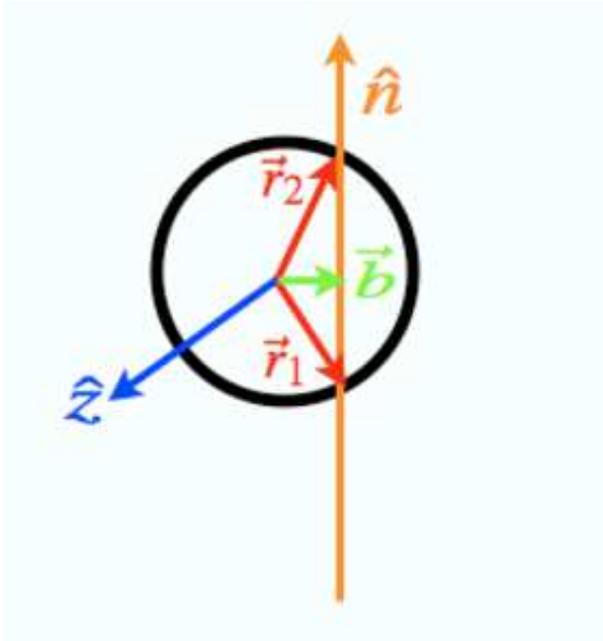}
\caption[]{Schematic illustration of the geometry for a single LBG
outflow. The black circle represents the shock front, the orange arrow
labelled $\hat{n}$ is the line-of-sight from the observer, the red arrows
labelled $\vec{r}_1, \vec{r}_2$ are the entry and exit points of the
line-of-sight on the outflow, the blue arrow labelled $\hat{z}$ is the
symmetry axis of the outflow and the green arrow labelled $\vec{b}$ is the
impact parameter of the line-of-sight.}
\label{fig:outflow} 
\end{figure}

First we consider the case where one line-of-sight intersects a single LBG
outflow.  Averaging over symmetry axis orientation and impact parameter,
the expectation value of the fractional temperature perturbation is \be
\label{eq:sing_los} \left \langle \left( \frac{\Delta T}{T} \right)_i
\right \rangle = \int d^2\hat{z} P(\hat{z}) \int d^2 \vec{b}~P(b)
\left(\frac{\Delta T}{T} \right)_i.  \ee Here $\hat{z}$ is the symmetry
axis of the outflow and $\vec{b}$ is the impact parameter at which the
line-of-sight enters the LBG (see Fig. \ref{fig:outflow} for definitions
of the variables we use in describing the LBG outflow).
We assume uniform probability distributions for the symmetry axis
orientation, $P(\hat{z}) = 1/4\pi$, and for the impact parameter,
$P(\vec{b}) = 1/\pi R^2.$

Now we change the integration variable from the impact parameter $\vec{b}$
to the location of entrance point of the line-of-sight into the LBG 
outflow $\hat{r}$. Thus, the expectation value defined in Eq. (\ref{eq:sing_los}) 
can be written as 
\be \left \langle \left( \frac{\Delta T}{T} \right)_i \right \rangle =
\int_{4\pi} \frac{d^2\hat{z}}{4\pi} \int_{2\pi} \frac{d^2\hat{r}}{\pi}
\left(\frac{\Delta T}{T} \right)_i, \ee 
where the integration over the
location of the impact parameter is restricted to a single hemisphere.
Performing this integration we find that the signal vanishes on average, as
expected from the fact that the net signal for a given LBG is just as
likely to be negative as it is to be positive.

There will be a non-zero contribution from a given LBG since the induced
anisotropies will be different at the entry and exit points of the
line-of-sight through the shock front. Because the LBG symmetry axes is
randomly oriented with respect to the direction of the observer, the signal
vanishes once the average over this direction is done. This implies that
the mean signal and therefore the one-point function vanishes. We are
ultimately interested in the two-point correlation function and the related
power spectrum. There are two contributions to the power spectrum
\citep{Cooray02}. The first is a clustering (two halo) term, originating
from the correlated perturbations in the cold dark matter density produced
during inflation.  The second is a Poisson (one halo) term, originating
from Poisson fluctuations in the number density of halos. The one-point
function vanishes because net temperature anisotropy produced by a given
LBG is uncorrelated with the signal from other LBGs along the
line-of-sight. The clustering term only implies that the number density of
LBGs nearby another LBG is greater than average, not that the symmetry axes
are somehow correlated\footnote{On small scales, the asphericity of the
outflows may be correlated because they propagate into the same
inhomgeneous IGM, or because tidal gravitational forces produced
correlations in the shapes of nearby galaxies \citep{Mackey02}.}. Since
there is no correlation in the signals between the two distinct
lines-of-sight, there will be no contribution to the resulting CMB power
spectrum from a clustering term.

\subsection{Two Sightlines}
For a single sightline through each outflow we found that the two-point
correlation function vanishes, as the symmetry axes of LBGs are randomly
oriented. When both lines-of-sight intersect the same LBG this cancellation
does not take place. The average value of the two-point temperature
anisotropy when both lines of sight intersect the same LBG is \be \left
\langle \left( \frac{\Delta T}{T} \right)_i^2 \right \rangle = \int
d^2\hat{z} P(\hat{z}) \int d^2 \vec{b}_1~P(\vec{b}_1) \int
d^2\vec{b}_2~P(\vec{b}_2) \left(\frac{\Delta T}{T} \right)^2_i \ee
Performing the relevant integrations we find a non-zero answer when the
product $n_e(\vec{r})~v(\vec{r})$ has odd parity. The induced fluctuations
produced when two lines of sight intersect the same LBG are due to Poisson
fluctuations. As mentioned above, we parameterize the deviation from
sphericity with a fudge factor $\epsilon$. Then the expectation value for
two lines-of-sight intersecting the same LBG outflow region is \be
\label{eq:dT2} \left \langle \left( \frac{\Delta T}{T} \right)_i^2 \right
\rangle = \sigma^2_T n^2_e \delta R^2 \frac{\epsilon^2 v^2}{c^2}.  \ee

\section{Poisson Fluctuations}
We have seen that there is an exact cancellation when the two
lines-of-sight intersect two different LBG outflows and that a non-zero
signal arises when the two lines-of-sight intersect a single LBG
outflow. Correlations in the one LBG terms 
are produced by Poisson fluctuations in the number of intercepted LBGs. We
will analyze this effect in two stages: first, we will consider the effect
along a given line-of-sight (pencil-beam survey) and then generalize to the
case of a finite beam size.

\subsection{Pencil Beam Survey}
The temperature anisotropies induced by LBGs of halo mass between $M$ and
$M+dM$, formed between redshifts $z_f$ and $z_f + dz_f$ and scattering the
CMB between redshifts $z$ to $z+dz$ is \be \frac{\Delta T}{T} =
\sum_{n=1}^{\infty} P_{dN}(n_{\rm LBG}) \sum_{i=1}^{n_{\rm LBG}}
\left(\frac{\Delta T}{T}\right)_i, \ee where $P_N(n_{\rm LBG})$ is the
Poisson probability that $n_{\rm LBG}$ LBGs are observed and $dN$ is the
mean number of LBGs in the redshift interval between $z$ and $z+dz$, \be dN
= \pi R^2 \frac{-c~dz}{(1+z)H(z)} \frac{d^2n}{dM dz_f} dM dz_f, \ee where
$R$ is the radius of an outflow at redshift $z$ produced by an LBG of mass $M$
formed at redshift $z_f$. The total signal is found by integrating over
$dM$, $dz_f$ and $dz$.  The expectation value toward a given line of sight
vanishes since the expectation value for the temperature anisotropy
produced by a single LBG vanishes. Nevertheless the variance of the signal
does not vanish due to Poisson fluctuations. For a single sightline through
an LBG outflow region the contribution to the temperature anisotropies can
be either positive or negative with equal probability, however for two
sightlines the contribution is always non-negative.  The resultant
variance, which is the case of two sightlines at a separation less than the
characteristic angular size of LBG outflows, is \be \left\langle
\left(\frac{\Delta T}{T}\right)^2\right\rangle = \int dN \left\langle
\left(\frac{\Delta T}{T}\right)_i^2\right\rangle.  \ee The {\it rms} value
of the anisotropy can be evaluated in terms of the unknowns and the time
changing Legendre coefficients $n_{\ell}$ and $v_{\ell}$. For simplicity,
we adopt the spherically symmetric solution for the shock radius and
velocity, and parameterize the degree of asymmetry in the outflow by the
fudge factor $\epsilon$.

\subsection{Window Function Effects}
The average angular size of an LBG outflow $\bar{\theta}_{\rm LBG} \approx
10\arcsec$ is below the resolution of the upcoming generation of
experiments, and so we must properly account for the beam's window
function. The observed temperature anisotropy will be an average over a
window function $W(\hat{n})$ \be \frac{\Delta \tilde{T}}{T}(\hat{n}) = \int
d^2 \hat{n}'~W(\hat{n}-\hat{n}')~\frac{\Delta T}{T}(\hat{n}').  \ee For
simplicity, we will take the window function shape to be a top hat of
angular size $\theta$, namely $W(\hat{n}) = 1/\theta^2 \mbox{ if }
|\hat{n}| \le \theta$, and $W(\hat{n})= 0 \mbox{ if } |\hat{n}| > \theta$.

The variance in an angular aperature defined by the window function, \be \label{eq:cmb_var}
\left \langle \left( \frac{\Delta \tilde{T}}{T}\right)^2
\right\rangle_{\theta} = \int d^2\hat{n}'_1
d^2\hat{n}'_2~W(\hat{n}'_1)~W(\hat{n}'_2)~\left\langle\frac{\Delta
T}{T}(\hat{n}'_1)~\frac{\Delta T}{T}(\hat{n}'_2) \right\rangle, \ee is
related to the power spectrum as \be \frac{\ell(\ell+1)C_{\ell}}{2\pi}
\approx \left \langle\left( \frac{\Delta \tilde{T}}{T}\right)^2
\right\rangle_{\theta = 2\pi/\ell}.  \ee

In order for Poisson fluctuations to give a nonzero result, the two lines
of sight $\hat{n}'_1$ and $\hat{n}'_2$ must intersect the same
LBG. Therefore they must be separated by less than $\bar{\theta}_{\rm
LBG}$. This requirement allows us to evaluate Eq. (\ref{eq:cmb_var}) as \be
\left \langle \left( \frac{\Delta \tilde{T}}{T}\right)^2
\right\rangle_{\theta} = \frac1{\theta^2} \int dN \left( \frac{R}{D_{\rm
A}(z)} \right)^2 \left \langle \left( \frac{\Delta T}{T}\right)_i^2
\right\rangle.  \ee Note that the Fourier multipole corresponding to
$\bar{\theta}_{\rm LBG}$ is $\ell = 2\pi/\bar{\theta}_{\rm LBG}$.  Here
$D_{\rm A}(z)$ is the angular diameter distance to redshift $z$.

A broad window function allows for a larger number of LBGs within the beam.
The window function is normalized such that it integrates to unity, and so
one is observing the fractional fluctuations in the signal as an average
over $\theta^2/\theta^2_{\rm LBG}$ independent coherence patches in the
beam. This is equivalent to a fractional fluctuation of $1/\sqrt{N}$ as
expected from Poisson fluctuations.

\section{Results}
We numerically solve for the evolution of the shock front in the thin shell
approximation \citep{Tegmark93, Furlanetto03}, taking into account the
effects of the halo gravity and the self gravity of the mass shell, the
internal pressure of shocked IGM and the acceleration due to the
cosmological constant. When calculating the internal pressure we include
Compton cooling, adiabatic cooling, and the addition of shock heated
gas. Initially we assume that a fraction $f_{*} = 0.1$ of the baryons
assembled into the central LBG form stars with a Scalo initial mass
function. In this case there is one supernova per $126$ \msun of star
formation \citep{Furlanetto03}. Supernovae characteristically produce $10^{51}$ 
ergs of energy, but only a
small fraction of this energy, $f_{\rm SN}$, couples to the outflow, with
the rest being radiated away. We adopt a value of $f_{\rm SN}=0.01$; see
\cite{Furlanetto03} and reference within for a more detailed description
of our outflow model.

Assuming a Sheth-Tormen mass function \citep{Sheth99} we include the
effects from all possible halos above the minimum galaxy mass. The lowest
galaxy mass is determined by the maximum between the Jeans filtering mass
and the cooling mass (dictated by the halo's ability to cool through atomic
hydrogen line emission)\footnote{Note that we include haloes with masses
below the observational sensitivity for LBGs, as there is no fundamental
reason to exclude these haloes.}. We allow the LBGs to form
between\footnote{We include this lower redshift limit because the
phenomenon of downsizing, as well as the evolution of AGN luminosity
function, implies that massive galaxies finished forming stars around that
redshift.} $z=2$ and $z=8$. We continue to allow the outflow to evolve
until its velocity equals to Hubble flow at its radius from the LBG or
until $z=0$. Increasing the upper redshift has little effect on our results
because of the higher minimum LBG mass, as well as, the higher velocity of
the Hubble flow which causes the LBG outflows to merge with the IGM at an
relatively earlier time. Decreasing the minimum LBG mass could have a
significant effect on our results because the ratio of the initial outflow
velocity to the halo escape velocity at the initial radius scales as
$v_{\rm init}/v_{esc} \propto M^{-2/9}$.  However, in low mass haloes the
star formation timescale increases and supernova feedback becomes capable
of decreasing $f_{*}$.  Some starburst activity continues to lower
redshifts, albeit with a reduced intensity compared to high redshift star
formation, changing the lower redshift of from $z_F = 2$ to $z_F = 1$ would
change our results by a factor of $2$. Since the observed star formation
efficiency decreases with decreasing $z_F$, our model, which assumes that
star formation and the subsequent supernova feedback only depends on the
LBG mass, overestimates this change.

\begin{figure}
\includegraphics[width=8cm]{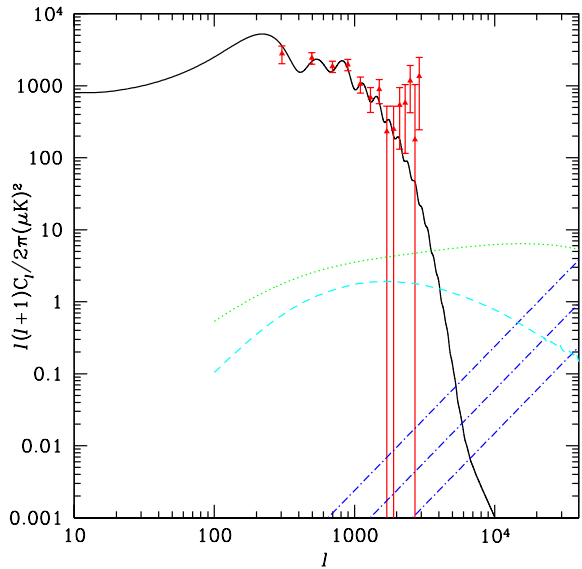}
\caption[]{Power spectra produced by several secondary mechanisms. The
contribution from the LBG outflow is shown as the three blue dot-dashed curves for
$\epsilon = 1, 0.5, 0.25$ from top to bottom. The primary CMB anisotropies are shown in the
solid-black curve; the Ostriker-Vishniac and patchy reionization
contributions as calculated by \cite{Mcquinn06} are dotted-green and
dashed-light-blue respectively. The observed CBI data points are plotted as
red triangles.}
\label{fig:power_spec} 
\end{figure}

In Fig. \ref{fig:power_spec} we show the CMB power spectrum produced by
several secondary mechanisms. The three blue dot-dashed curves denote the LBG outflow
signal calculated in this {\it Letter} for the values of $\epsilon = 1,
0.5, 0.25$ from top to bottom. The green dotted curve delineates the Ostriker-Vishniac 
effect and the dashed cyan curve describes the patchy reionization effect as 
calculated by \cite{Mcquinn06}\footnote{Note that \cite{Mcquinn06} used a 
different cosmological model with a higher value of $\sigma_8 = 0.9$. This 
artificially raises the amplitude of the results compared to value of $\sigma_8 = 0.73$
adopted in this work in line with WMAP3.}  
Also, for comparison, we show the primary CMB anisotropies (solid, black curve) and
the small scale Cosmic Background Imager (CBI) data points (red triangles) 
\citep{Readhead04}.

To characterize our results, let us mention some typical quantities for a
common LBG halo of mass $5 \times10^{9} M_{\odot}$ formed at $z_F = 2$. In
this case the outflow reaches a maximum comoving radius of 150 \kpc ~ at
$z= 0.9$ before it merges with the Hubble flow. At late times the outflow
velocity with respect to the local Hubble flow is $(v-HR)/c \sim 7 \times
10^{-4}$ and the Thomson scattering optical depth through the shock front
is $\tau \approx 10^{-5}$.

These small characteristic values for the Thomson scattering optical depth
and outflow velocity imply that the signal should not be notably
polarized. Thomson scattering of a radiation field containing a quadrupole
moment will produce polarized radiation.  There are two standard ways in
which scattering by a halo can produce polarization -- (i) photons will
double scatter in the halo; (ii) the peculiar velocity of the scatterer
will induce a quadrupole moment in the radiation field
\citep{Zeldovich80}. In the first case, the radiation can scatter in the
halo producing an anisotropic radiation field; a second scattering of that
radiation field can produce polarization at the level $\mathcal{O}(\tau^2
v/c)$.  In the second case, the peculiar velocity of the scatterer
perpendicular to its line-of-sight with respect to the observer will induce
a quadrupole in the incident radiation field at the order
$\mathcal{O}(v^2/c^2)$.  A fraction $\tau$ of the radiation field will
scatter and become polarized at the level $\mathcal{O}(\tau v^2/c^2)$.
Since values of $\tau$ and $v/c$ are so small, we conclude that the
polarization power spectrum (which is sixth order in the small parameters
of $\tau$ and $v/c$) is negligible.

\section{Discussion}

The contribution of outflows from Lyman-break galaxies to the CMB
power-spectrum on arcminute scales is proportional to the square of their
characteristic level of deviation from sphericity, $\epsilon^2$. Future CMB
experiments could therefore calibrate the intricate feedback process of
galactic outflows on the IGM. Most tools used to study these feedback
processes focus on the inner few \kpc ~of the LBGs, even though the shock
front is typically located at several of tens or hundreds of \kpc. The
secondary CMB anistropies calculated in this work provide a unique probe of
these extended perturbed regions around starburst galaxies at high
redshifts.

Even though the amplitude of the power spectrum produced by this effect is
small, the signal has distinctive spectral and spatial characteristics. The
power spectrum has the same scaling with Fourier multipole ($\propto
\ell^2$) as radio or IR point sources. However, the frequency dependence of
the anisotropies produced by our effect has the standard blackbody
spectrum, whereas the radio point sources have a power law frequency
spectrum. This distinct feature of our effect may allow future small scale
experiments, many of which have excellent frequency coverage, to separate
the anisotropies produced by LBGs from radio point sources.

\bigskip

{\bf Acknowledgements}

This work was supported in part by Harvard university grants. D.B. thanks
the Harvard Institute for Theory and Computation for its hospitality when
this work was completed, and acknowledges helpful conversations with Re'em
Sari and Niayesh Afshordi. We also thank Matt McQuinn for making available
the numerical results from his work.

 \end{document}